\documentclass[10pt,article]{article}
\usepackage{opex3}
\usepackage{color}
\usepackage{cite}
\begin{document}

\title{Tunable, Stable Source of Femtosecond Pulses Near 2~\textmu m via Supercontinuum of an Erbium Mode-Locked Laser}

\author{Andrew~Klose$^{1,*}$, Gabriel~Ycas$^{1}$, Daniel~L.~Maser$^{1,2}$, and~Scott~A.~Diddams$^{1,2}$}

\address{$^1$National Institute of Standards and Technology, Boulder, CO\\
$^2$Department of Physics, University of Colorado, Boulder, CO}

\email{$^*$andrew.klose@nist.gov} 



\begin{abstract}
A source of ultrashort pulses of light in the 2~\textmu m region was constructed using supercontinuum broadening from an erbium mode-locked laser. The output spectrum spanned $1000$~nm~to~$2200$~nm with an average power of 250~mW.  A pulse width of 39~fs for part of the spectrum in the 2000~nm region, corresponding to less than six optical cycles, was achieved.  A heterodyne measurement of the free-running mode-locked laser with a narrow-linewidth continuous wave laser resulted in a near shot noise-limited beat note with a signal-to-noise ratio of 45~dB in a 10~kHz resolution bandwidth.  The relative intensity noise of the broadband system was investigated over the entire supercontinuum, and the integrated relative intensity noise of the 2000~nm portion of the spectrum was $1.7\times10^{-3}$.  The long-term stability of the system was characterized, and intensity fluctuations in the spectrum were found to be highly correlated throughout the supercontinuum.  Spectroscopic limitations due to the laser noise characteristics are discussed.
\end{abstract}


\bibliography{refs}
\bibliographystyle{osajnl} 


\section{Introduction}

Optical frequency combs based on femtosecond, mode-locked, fiber lasers have been exploited for many applications including precision spectroscopic measurements using a variety of detection techniques \cite{diddams07, keilmann04, coddington08, giaccari08, mandon09, ideguchi13}.  Lasing in the near-IR region of 1~to~2 \textmu m has been achieved using optical fibers doped with various rare earth elements, including Yb, Er, and Tm.  The use of nonlinear optical processes, including supercontinuum generation with highly nonlinear optical fiber, optical parametric oscillation, and difference frequency generation, has lead to the development of frequency combs throughout the near-IR and mid-IR spectral regions \cite{ycas12, kumkar12, hartl12, adler12, hoogland13, zhu13, neely11, schliesser12}.  In the time domain, the short pulses available from coherent mode-locked lasers in the near-IR have advantages for microscopy \cite{xu13}, cavity enhancement for driving harmonic generation \cite{cingoz11}, and creating ultrashort pulses in the mid-IR with stable carrier-envelope phase \cite{erny09,chalus09}. Therefore, significant effort has been devoted to generate NIR/MIR frequency combs with broad bandwidths and near transform-limited pulse widths \cite{adler12, hoogland13, col14}. Along these lines, pulses as short as two optical cycles have been realized \cite{sell09}.

For the envisioned application of frequency comb spectroscopy, the source noise characteristics, both short-term relative intensity noise and long-term spectral stability, affect the ultimate detection sensitivity.  In particular, low-noise sources with stable output spectra are advantageous for use in experiments outside of the controlled environment of a scientific laboratory.  One undesirable property of fiber systems constructed with standard single mode fibers is polarization rotation resulting from mechanical stress on the fiber and/or changes in ambient conditions.  Such behavior requires frequent tuning of optical components, and leads to performance which may not be readily reproducible.  This limited stability is a significant drawback for the development of a fieldable instrument. One method to overcome these difficulties and increase the robustness of fiber sources is the implementation of polarization maintaining (PM) optical fibers. Indeed, the recent development of a PM source has shown quite remarkable performance characteristics \cite{sinclair13}. 

The 2000~nm to 2500~nm spectral region is of interest for atmospheric spectroscopic measurements, as relatively strong greenhouse gas absorption features, in combination with the transparency of water, provide a window for precision measurements of CO$_2$ and CH$_4$.  A few efforts have focused on the use of Tm-doped optical fibers to generate comb sources in this spectral region \cite{hartl12,adler12,hoogland13}.  However, the cost-effective and technologically mature Er:fiber frequency comb, used in combination with supercontinuum generation, has practical advantages for 2~\textmu m generation that make it worth continued investigation. Here, a highly-stable Er:fiber-based supercontinuum frequency comb with a tunable output spectrum and sub-40~fs pulses in the 2 \textmu m region is presented. A rigorous characterization of this source in terms of frequency noise, short- and long-term amplitude fluctuations and spectrally-correlated noise is provided.  Limitations related to spectroscopic measurements are discussed.

\section{Experimental Setup}

\begin{figure}
\centering
\includegraphics{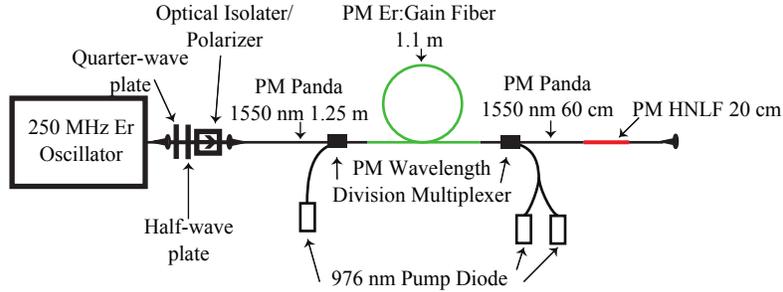}
\caption{Schematic of the PM supercontinuum source discussed in the presented work (note: not drawn to scale).}
\label{fig:schematic}
\end{figure}

A schematic of the laser system is presented in Figure~\ref{fig:schematic}.  The system was built upon a commercial 250~MHz mode-locked Er:fiber laser which had a pulse width of approximately 100~fs.  In future work, we envision that this oscillator could be replaced by an all-PM mode-locked Er:fiber laser \cite{sinclair13}, but at present linearly-polarized output was created using a combination of waveplates and a polarizer that was part of an optical isolator. Approximately 25~mW of linearly-polarized light was launched into a PM Er:fiber amplifier.  The pulses were stretched in anomalous dispersion single-mode fiber such that the group-delay dispersion of the pulses was near $-40,000$~fs$^2$ entering the Er-doped fiber.  The gain fiber was normally-dispersive at 1550~nm ($D=-25.2$~ps/nm/km), had a mode field diameter of 6.5~\textmu m at 1550~nm, and small signal absorption of 80~dB/m at 1530~nm. The amplifier was pumped with three 976~nm diodes, configured with a single forward pump and two reverse pumps.  The maximum pump power delivered to the gain fiber was 750~mW (1300~mW) in the forward (reverse) direction.  An average output power of 350~mW was achieved after amplification.  The amplified pulses were compressed to $<90$~fs in PM single-mode fiber and launched through 0.2~m of PM highly nonlinear fiber (HNLF) ($D$~$=$~+1.9~ps/nm/km, ZDW~=~1490~nm, $\gamma$~$\sim$~20/W/m) after which 250~mW of light spanning $1.0$~to~$2.2$~\textmu m emerged.

\begin{figure}
\centering
\includegraphics{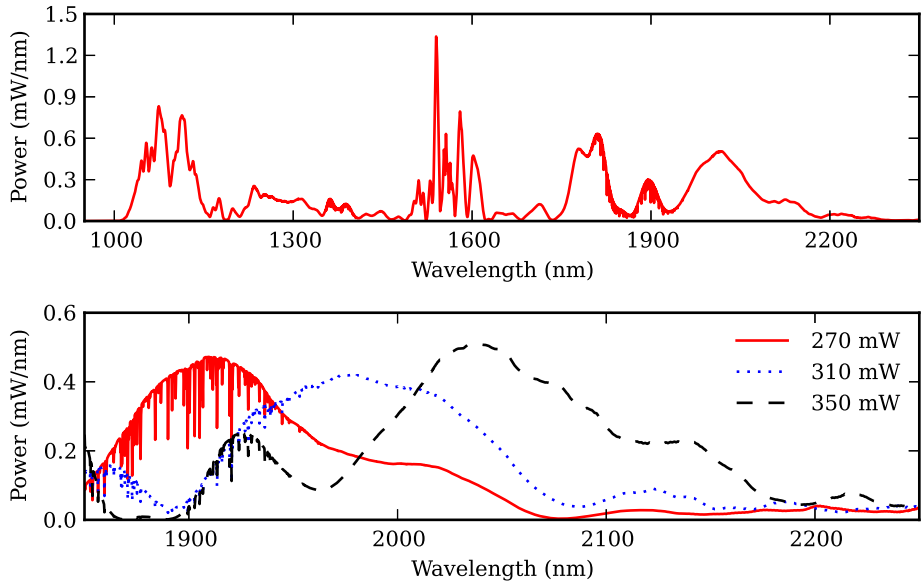}
\caption{Top: Measured spectrum after the HNLF. Bottom: Spectral output near 2~\textmu m for varying Er amplifier average output powers, denoted on the legend. The depletions near 1.9~\textmu m are due to water absorption in the OSA.}
\label{fig:spectra}
\end{figure}

A spectrum after the HNLF, measured using an optical spectrum analyzer (OSA), is presented in the top panel of Figure~\ref{fig:spectra}.  The portion of the spectrum with $\lambda > 1.85$~\textmu m contains over 30$\,$\% (75~mW) of the optical power in the supercontinuum.  Additionally, the relative location of the Raman-shifted spectral peaks near 1.9~\textmu m and 2.0~\textmu m were deterministically tuned by altering the average power launched into the HNLF.  This was accomplished by varying the Er amplifier 976~nm pump power from 1.2~W to 2.0~W, corresponding to average output powers of 270~mW to 350~mW launched into the HNLF.  The $\lambda > 1.85$~\textmu m portion of the supercontinuum spectra obtained as a function of 976~nm pump power tuning is shown in the bottom panel of Figure~\ref{fig:spectra}.

\begin{figure}
  \centering
  \includegraphics{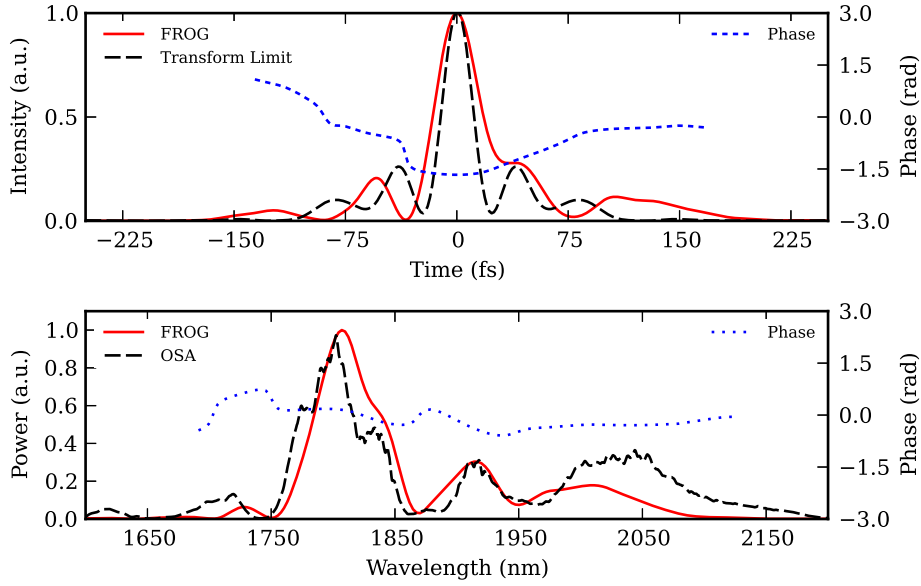}
\caption{Top: Pulse retrieved from SHG FROG measurement of compressed pulse. Bottom: Retrieved FROG and OSA-measured spectra of compressed pulses.}
\label{fig:frog}
\end{figure}

The temporal characteristics of the 2~\textmu m portion of supercontinuum were investigated using second harmonic generation frequency-resolved optical gating (SHG FROG) \cite{trebino93}.  The 2~\textmu m portion of the pulses emerging from the free-space coupled output of the HNLF was determined to have a pulse width of 90~fs.  A silicon prism pair was used to compress the pulses to 39~fs, significantly shorter than 2~\textmu m pulses generated or amplified with Tm-doped fiber \cite{adler12,hoogland13,leindecker12}.  The retrieved pulse and spectrum obtained from a FROG measurement of the compressed pulses are displayed as solid lines in Figure~\ref{fig:frog}, respectively.  The retrieved temporal phase and spectral phase of the pulse are depicted with dotted lines in Figure~\ref{fig:frog}.  The retrieved FROG spectrum, and a direct OSA measurement of the spectrum after the prism compressor, are presented as a solid and dashed lines, respectively, in the bottom panel in Figure~\ref{fig:frog}. The discrepancy between the FROG-retrieved and OSA spectra is attributed to the limited phase matching bandwidth of the nonlinear crystal used in the SHG FROG apparatus. The temporal Fourier transform limits, calculated from the OSA and FROG-retrieved spectra, were found to be 22~fs and 23~fs, respectively.  The transform-limited pulse width deduced from the OSA spectrum is plotted as a dashed line in the top panel of Figure~\ref{fig:frog}.

\section{Source Coherence and Noise Characteristics}

\begin{figure}
\centering
\includegraphics{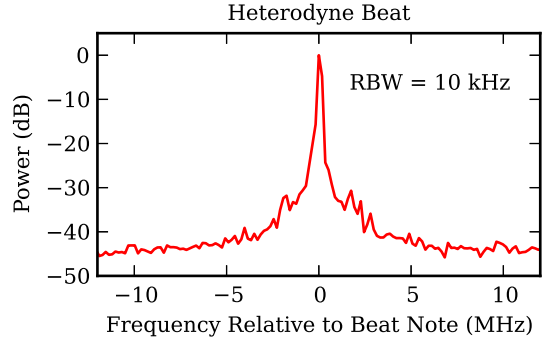}
\caption{Heterodyne beat note of doubled supercontinuum frequency comb with a single frequency laser near 990~nm.}
\label{fig:beat-note}
\end{figure}

Phase coherence of the 2~\textmu m portion of the output spectrum was investigated by conducting a free-running heterodyne measurement of the supercontinuum with a single-frequency external-cavity diode laser.  The supercontinuum near 1980~nm was frequency-doubled using a suitable periodically-poled lithium niobate crystal and beat against the continuous-wave (CW) laser near 990~nm on a fast photodiode. A representative RF spectrum of the photodiode signal, measured with a resolution bandwidth (RBW) of 10~kHz and averaged over 1.1~s, is presented in Figure~\ref{fig:beat-note}.  The heterodyne beat note between the CW laser and the nearest comb mode was observed to have a signal-to-noise ratio (SNR) of 45~dB, near the limit of 50~dB imposed by the power of the lasers and the shot-noise of the CW laser.

\begin{figure}
  \centering
  \includegraphics{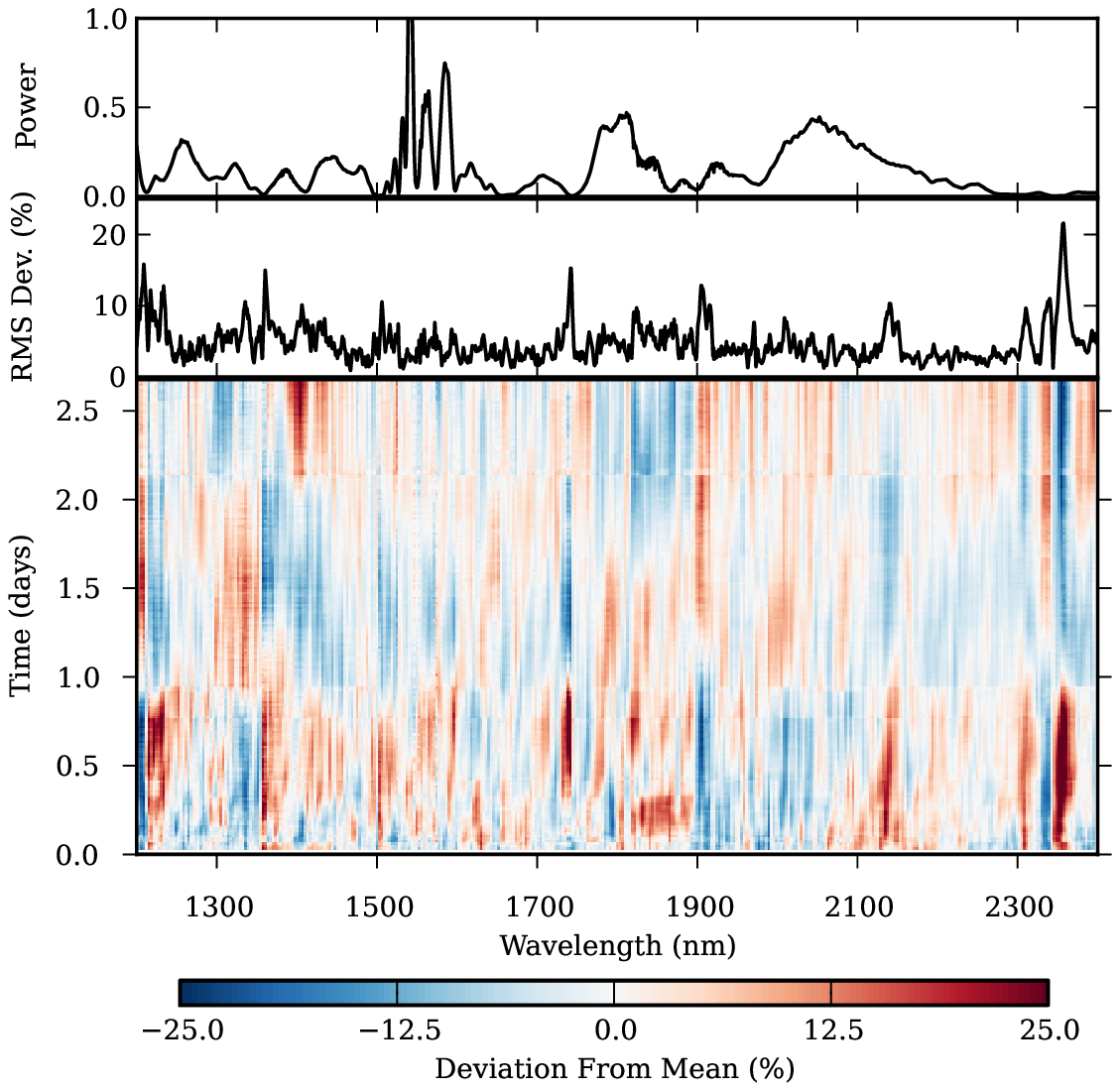}
\caption{Top: mean optical spectrum over 2.5~days of measurement.  Middle: RMS deviation from the mean of each resolution element.  Bottom: Contour plot of intensity variation of the source spectrum over 2.5 days.  The percent deviation from the mean intensity value for each resolution element of the spectrometer is depicted by the color contour.}
\label{fig:waterfall}
\end{figure}

\begin{figure}
  \centering
  \includegraphics{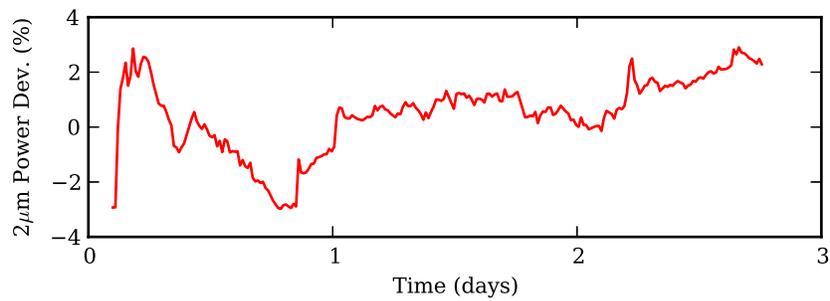}
\caption{RMS deviation of the integrated 2~\textmu m peak power from the mean value over 2.5~days.}
\label{fig:long-term-stab}
\end{figure}

The long-term stability of the system was studied by measuring the output spectrum of the system every 15 minutes over 2.5 days.  A contour plot depicting the variation of the optical spectrum from the mean over the 2.5 days is presented in Figure~\ref{fig:waterfall}.  Many of the large variations occur at wavelengths where there is a dearth of optical power, for example, near 1350~nm, 1500~nm, and 1750~nm.  However, sizable fluctuations at the 15$\,$\% level were observed at wavelengths with significant optical power. The rms deviation of the output power over all spectrum measurements, determined for each resolution element, is displayed in the middle panel of Figure~\ref{fig:waterfall}. Additionally, the rms deviation of the integrated optical power of the $1.95$~\textmu m~to~$2.20$~\textmu m spectral peak from the mean value was calculated for each spectrum measurement; the result is presented in Figure~\ref{fig:long-term-stab}.  The rms deviation of the 2~\textmu m peak power was 1.3$\,$\%.

\begin{figure}
  \centering
  \includegraphics{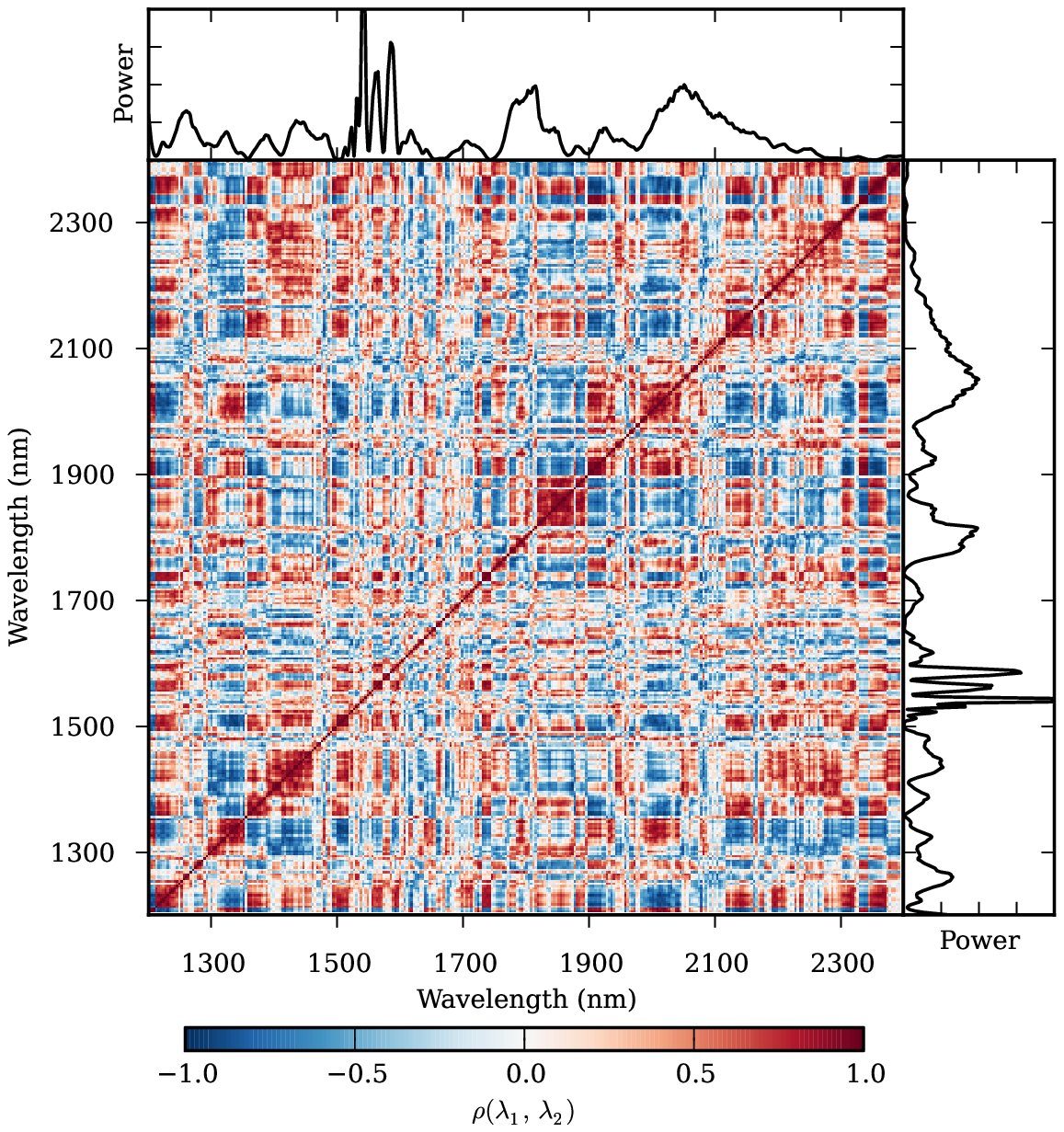}
\caption{Wavelength-dependent correlations of time-dependent spectral intensity fluctuations.  The mean optical spectrum is plotted on a linear scale above and to the right of the correlation contour plot.}
\label{fig:corr}
\end{figure}

Correlations in the long-term fluctuations of the time-dependent spectral measurements were investigated by calculating a correlation coefficient, $\rho$, for all pairs of wavelengths, $\lambda_1$, $\lambda_2$, as
\begin{equation}
\rho(\lambda_1,\lambda_2) = \frac{\big\langle I(\lambda_1)I(\lambda_2) \big\rangle - \big\langle I(\lambda_1)\big\rangle \big\langle I(\lambda_2) \big\rangle}
{\sqrt{\Big( \big\langle I^2(\lambda_1)\big\rangle-\big\langle I(\lambda_1) \big\rangle^2 \Big) \Big( \big\langle I^2(\lambda_2) \big\rangle-\big\langle I(\lambda_2) \big\rangle^2 \Big)}}.
\end{equation}
Here, $\rho(\lambda_1,\lambda_2)$ is the correlation coefficient of $\lambda_1$ and $\lambda_2$ and spans $-1 < \rho(\lambda_1,\lambda_2) < 1 $, where a value of $+1 (-1)$  is indicative of fully correlated (anti-correlated) fluctuations. Additionally, $I(\lambda_i)$ is the intensity of the optical spectrum at wavelength $\lambda_i$, and the angle brackets indicate averaged values over the 2.5 days of measurement.  Previously, this correlation technique has been used to investigate shot-to-shot variation of supercontinua generated from ultrafast lasers \cite{godin13,majus13}. However, the supercontinuum presented here is much broader than previous measurements, and the spectral fluctuations are investigated over a time scale of days, where an individual measurement of the spectrum took around 10~s.  A contour plot of the correlations of the intensity fluctuations is presented in Figure~\ref{fig:corr}.  The long-term wavelength-dependent intensity fluctuations are highly correlated across the supercontinuum. The average magnitude of off-diagonal correlation elements $(\lambda_1 \ne \lambda_2)$ was found to be $\big\langle | \rho(\lambda_1,\lambda_2) | \big\rangle =0.39$.  Qualitative features of the correlation plot match the supercontinuum spectral behavior.  For example, intensity fluctuations in the $2000$~nm~to~$2050$~nm portion of the spectrum are self-correlated, but anti-correlated with fluctuations near $2100$~nm~to~$2200$~nm, consistent with the spectral tuning as a function of input power to the HNLF.  As shown in bottom panel of Figure~\ref{fig:spectra}, as the input power increases the Raman-shifted peak near 2000~nm moves to longer wavelengths.  This implies that the red portion of the peak gains intensity, and the blue portion of the peak loses intensity.  Therefore, the blue and red sides of the peak have anti-correlated intensity noise.

\begin{figure}
\centering
\includegraphics{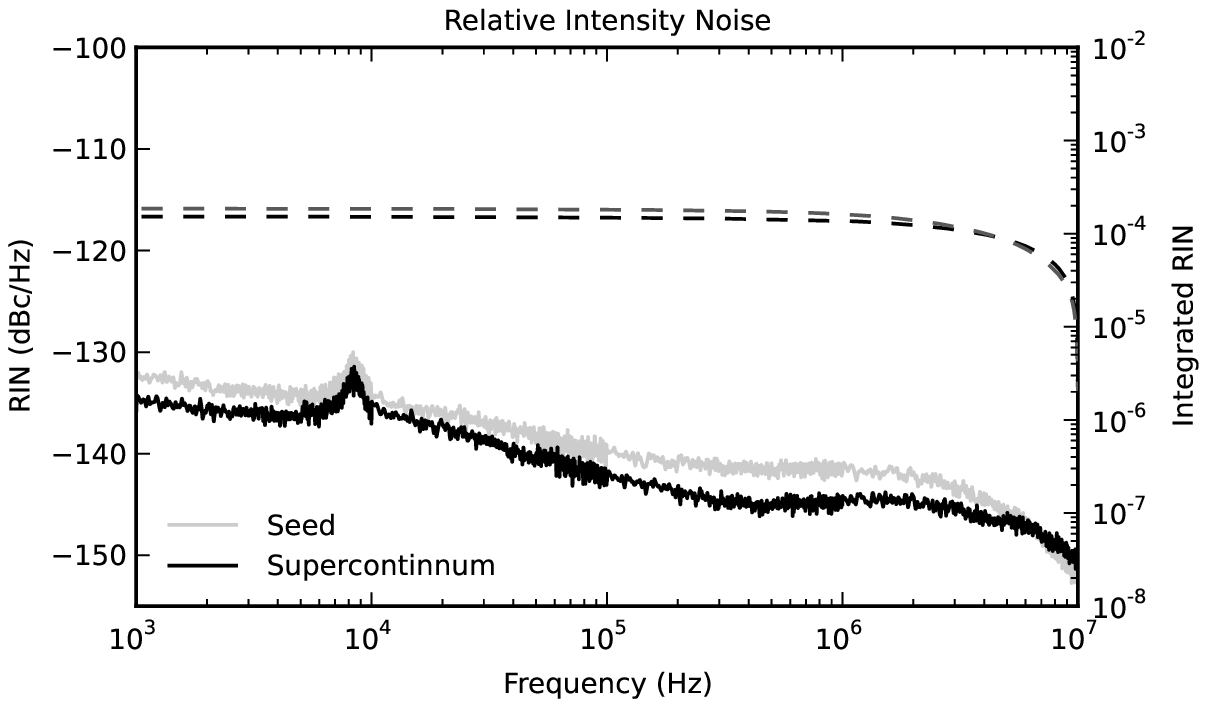}
\caption{RIN of seed source, in solid grey, and supercontinuum, in solid black.  The integrated RIN from 10~MHz, shown on the right-hand axis, is depicted with dashed lines.}
\label{fig:rin-seed}
\end{figure}

\begin{figure}
\centering
\includegraphics{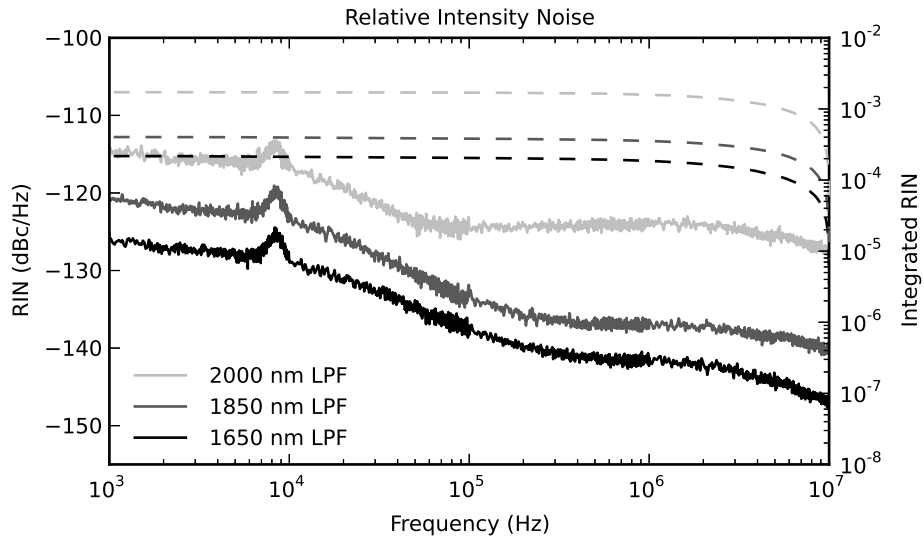}
\caption{RIN of long-pass-filtered portions of the supercontinuum.  The integrated RIN from 10~MHz, shown on the right-hand axis, is depicted with dashed lines.}
\label{fig:rin-lpf}
\end{figure}

The fast noise of the source was studied by performing relative intensity noise (RIN) measurements of the seed and supercontinuum. The detector used for these RIN measurements was sensitive from 1.2~\textmu m to 2.6~\textmu m. The RIN of the seed and the supercontinuum are plotted as solid lines in Figure~\ref{fig:rin-seed}.  The integrated RIN values from 10~MHz to 1~kHz are plotted as dashed lines in Figure~\ref{fig:rin-seed}. RIN measurements of filtered portions of the supercontinuum were carried out.  In particular, 1650~nm, 1850~nm, and 2000~nm long-pass filters were used to measure the RIN of these bandwidths of the supercontinuum, and the results are presented in Figure~\ref{fig:rin-lpf}. 

\begin{figure}
\centering
\includegraphics{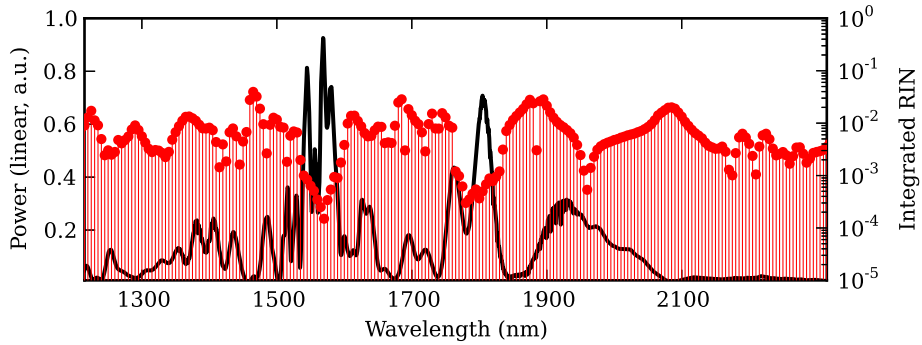}
\caption{Integrated RIN of 5~nm FWHM portions of the supercontinuum.  The integrated RIN values are shown as circles overlaid on the optical spectrum, which is plotted as a solid line.}
\label{fig:rin-mono}
\end{figure}

The integrated RIN value of $1.5\times10^{-4}$ of the full supercontinuum is similar to that of the seed source and consistent with previous Er-based supercontinuum results \cite{hori04}.  Moreover, the integrated RIN of the 2~\textmu m portion of the spectrum, $1.7\times10^{-3}$, is lower than ultrafast 2~\textmu m pulses amplified with Tm fibers where a result of $2.3\times10^{-3}$ over an integrated bandwidth of 500~kHz to 1~kHz was achieved \cite{hoogland13}. The integrated RIN of the 2000~nm portion of the present spectrum is $5.2\times10^{-4}$ when integrated from 500~kHz to 1~kHz. The increasing integrated RIN with decreasing measured optical bandwidth is indicative of anti-correlated intensity noise across the spectrum.  As the number of measured comb teeth decreases, the effect of averaging amplitude fluctuations of the anti-correlated portions of the spectrum leads to increased RIN.  To further characterize the intensity noise of the source, a monochromator was employed to filter the supercontinuum before photodetection.  The integrated RIN of 5~nm FWHM portions of the supercontinuum is plotted as a function of wavelength in Figure~\ref{fig:rin-mono}.  The integrated RIN of the narrow-bandwidth portions of the spectrum are within two decades across the supercontinuum, consistent with previous results of amplitude fluctuations across a fiber-based supercontinuum~\cite{ruehl11}.

The noise characteristics and stability of the source may be the limiting factor of spectroscopic measurements; therefore, it is important to consider these qualities in the context of implementing the source with a spectrometer.  In the current case, correlations of intensity fluctuation in the 2~\textmu m region of the spectrum indicate that intensity fluctuations are not independent comb tooth fluctuations, but rather correlated/anti-correlated fluctuations of the spectral envelope.  The long-term deviation of the 2~\textmu m peak power is at the percent level, and such fluctuations could be accounted for with a modest number of spectral background measurements during real-time data acquisition. Furthermore, it is important to consider fluctuations on the time scale of the spectrometer detector integration.  Currently, the most rapid detection and acquisition techniques are in the regime of 1~\textmu s to 1~ms, thus noise on this time scale must be minimized. Newbury and colleagues have investigated technical noise-limited SNR of dual-comb spectroscopy \cite{newbury10}, including the limitation introduced by laser RIN.  Within this framework, the additive technical noise is introduced as a small perturbation to the signal response, and a quality factor, $QF$, given as the product of the technical noise-limited SNR per unit time and number of resolved spectral elements, can be calculated.  The high-frequency RIN value of approximately $-125$~dBc/Hz of the 2~\textmu m portion of the supercontinuum would imply that a RIN-limited $QF$ of $1.0\times10^{6}$~Hz$^{1/2}$ would be achievable in a dual-comb spectroscopic experiment using a single balanced photodetector.  A typical RIN-limted $QF$ for fiber-laser systems lies in the range of $10^{6}$~Hz$^{1/2}$ to $10^{7}$~Hz$^{1/2}$ \cite{newbury10}.  The 2~\textmu m light generated by current system is at the edge of a supercontinuum, where the RIN is higher compared to the RIN of the seed source. Nonetheless, the RIN-limited $QF$ of the source presented here is still on the lower end of the typical range expected with mode-locked fiber lasers.   In the context of the source integrated into a spectroscopic system, the laser RIN may be the limiting noise contribution at a detected optical power higher than roughly 50~\textmu W.  At lower power, the system would limited by either detector or shot noise.

\section{Summary}
A tunable PM Er:fiber-based frequency comb in the 2~\textmu m region was developed via supercontinuum generation from the amplified output of a Er mode-locked laser. A prism pair was used to compress the 2~\textmu m portion of the supercontinuum to 39~fs.  The coherence of the 2~\textmu m portion of the comb was investigated by a free-running heterodyne measurement with a single-frequency laser, and a 45~dB SNR beat note was observed at a resolution bandwidth of 10~kHz.  Noise characteristics, including source RIN and long-term spectral stability, were investigated.  The results of the RIN measurements indicated that the integrated high-frequency noise of the 2~\textmu m portion of the source were 0.17$\,$\%. Furthermore, the total output power of the system was stable at the level of a few percent over a time scale of days. Spectroscopic measurements of atmospheric gases near 2~\textmu m, including carbon dioxide, will be pursued using the source presented in this work. While 2~\textmu m oscillator sources exist \cite{hartl12}, there are excellent all-PM oscillators at 1.5~\textmu m \cite{sinclair13} that, when coupled with the present PM broadening, could lead to a very attractive, fully-integrated, all-PM, completely fiber-based 2~\textmu m source.

\section{Acknowledgments}
The present work was funded in part by the NIST Greenhouse Gas and Climate Science Measurements Program.  The authors would like to thank T.A.~Johnson, D.~Hackett, F.C.~Cruz, I.~Coddington, and N.~Newbury for useful discussions, and M. Hirano of Sumitomo for providing the nonlinear fiber. A.K. acknowledges support from the National Research Council Postdoctoral Fellowship Program. This work is a contribution of the United States Government and is not subject to copyright in the United States.
\end{document}